\documentclass{article}
\usepackage{arxiv}
\usepackage{graphicx}
\usepackage[utf8]{inputenc} % allow utf-8 input
\usepackage[T1]{fontenc}    % use 8-bit T1 fonts
\usepackage{hyperref}       % hyperlinks
\usepackage{url}            % simple URL typesetting
\usepackage{booktabs}       % professional-quality tables
\usepackage{amsfonts}       % blackboard math symbols
\usepackage{nicefrac}       % compact symbols for 1/2, etc.
\usepackage{microtype}      % microtypography
\usepackage{lipsum}
\usepackage[marginal]{footmisc}

\title{Neighborhood-Enhanced and Time-Aware Model for Session based  Recommendation}

\author{
  Yang Lv\\
  University of Science and Technology of China\\
   \And
  Liangsheng Zhuang\\
  University of Science and Technology of China\\
   \And
  Pengyu Luo\\
  Hefei University of Technology\\
}

\begin{document}
\maketitle
\footnote{\noindent \textbf{Work in progress}}.

\begin{abstract}
Session based recommendation has become one of the research hotpots in the field of recommendation systems due to its highly practical value.Previous deep learning methods mostly focus on the sequential characteristics within the current session,and neglect the context similarity and temporal similarity between sessions which contain abundant collaborative information.In this paper,we 
propose a novel neural networks framework,namely Neighborhood Enhanced and Time Aware Recommendation Machine(NETA) for session based recommendation. Firstly,we introduce an efficient neighborhood retrieve mechanism to find out similar sessions which includes collaborative information.Then we design a guided attention with time-aware mechanism to extract collaborative representation from neighborhood sessions.Especially,temporal recency between sessions is considered separately.Finally, we design a simple co-attention mechanism to determine the importance of complementary collaborative representation when predicting the next item.Extensive experiments conducted on two real-world datasets demonstrate the effectiveness of our proposed model.
\end{abstract}

% keywords can be removed

\section{Introduction}

Recommender Systems(RS) are critical for online users to alleviate the problem of information overload in the era of big data and are widely used in various scenarios.Session based recommendation(SRS) has become one of the research hotpots in the field of recommendation systems due to its highly practical value.In common situations,user's profiles or past interactions may be not available for recommendation systems,since some users are anonymous/first-time visitors or  the online platform only tracks the session's identifier\cite{hidasi2015session}.To address this problem,session based recommendation is proposed to model limited interactions during the ongoing session while general recommendation  methods rely on user's profiles\cite{hidasi2015session}\cite{li2017neural}.

Previous works have made great progress for session based recommendation in the past few years.Early works are devoted to discovery item-to-item relations,such as transition relation and co-occurrence relation. Markov Chain(MC)\cite{garcin2013MC}\cite{he2009MC} and ItemKNN\cite{davidson2010ItemKNN}\cite{linden2003ItemKNN} are typical examples. Markov Chain assumes the next action is based on the previous one and learns the transition patterns between items.ItemKNN computes the similarity between items based on their co-occurrence frequency. However, these item-to-item models lack the ability of learning complex dependencies within the current session.
Recent studies  utilize neural network to model the whole action sequence.For instance,Hidasi et al.\cite{hidasi2015session} apply recurrent neural networks(RNN) for session based recommendation and treat this problem as time series prediction.Li et al.\cite{li2017neural} improve the RNN-based model by proposing an attention mechanism  to capture the user's main purpose additionally.Liu et al.\cite{liu2018stamp} highlight the importance of short-term memory by employing an attentive network to model the last item separately.

It is noteworthy that the aforementioned deep learning methods only focus on sequence modeling and are based on the limited actions of the current sessions, while rich information of collaborative filtering still has the potential to be further exploited.A well-known
intuition is that similar sessions tend to click similar items.SessionKNN\cite{jannach2017knnrnn} compares the entire current session with past sessions to determine the items to be recommended based on their occurences in k most similar past sessions(neighbors).But SessionKNN ignores the user's sequential behavior and interest shift within the current session,which is extremely essential because the orders of the clicked items  indicate user's main intention.Recently,CSRM
\cite{wang2019collaborative} is proposed to retrieve neighborhood sessions by applying an inner memory encoder(to remember recent sessions' representations) and an outer memory encoder(to identify neighbors).However,the CSRM model lacks an efficient retrieval capability and can't associate real similar neighborhood sessions due to memory limitation (the memory network only remembers the last 1w sessions refers to the implemented code for example).

\begin{figure}[ht] 
  \setlength{\abovecaptionskip}{-0.2cm}  
  \setlength{\belowcaptionskip}{0.1cm} 
    \centering %使图片居中显示
    \includegraphics[width=1\textwidth]{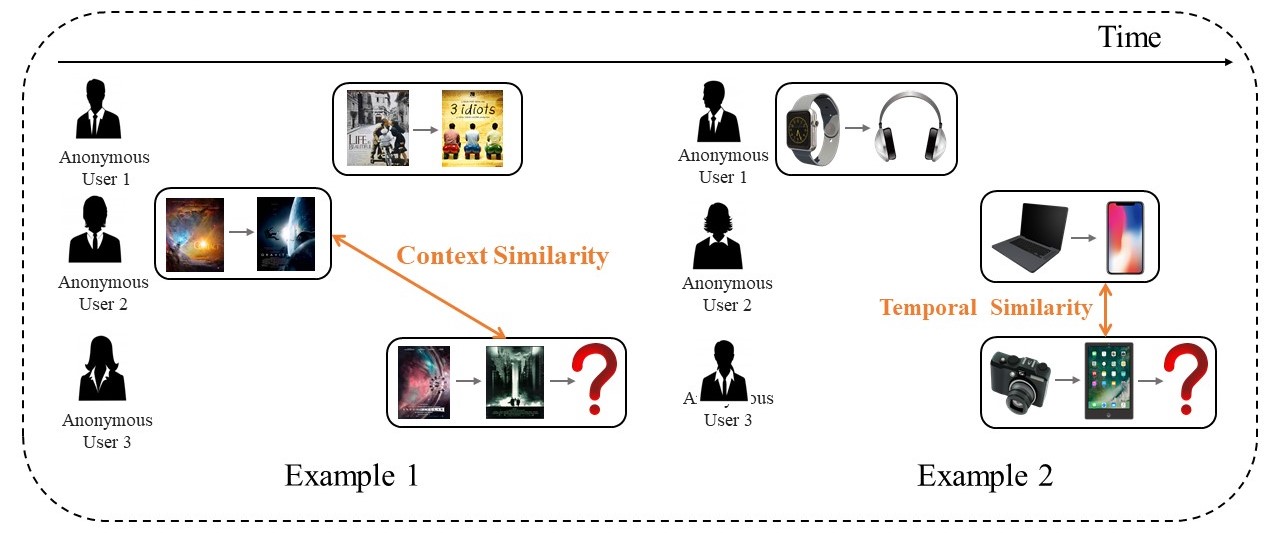} 
    \caption{Similarity:Context similarity and temporal similarity between sessions. Example 1 shows context similarity which provides collaborative information for recommendation, i.e. other science fiction movies. Example 2 shows recency similarity which provides hot and seasonal information for recommendation, i.e. the latest version of the iPhone.}
    \label{Similarity} 
\end{figure}

To tackle the above problems,we propose a novel neural networks framework,namely Neighborhood Enhanced and Time Aware Recommendation Machine(NETA) for session based recommendation.We argue that the similarity between sessions mainly exists in two aspects, namely context similarity and temporal similarity, as shown in detail in Figure 1. We consider the task of SRS from the perspective of the combination of neighborhood model and sequential model.Figure 2 illustrates the workflow of the proposed model. Firstly,we introduce an efficient neighborhood retrieve mechanism to find out k most similar sessions for the current session,which shares similar behavior pattern with the current session and includes collaborative information that indicates the next item.Then,in order to extract complementary representation(i.e.collaborative information) from neighborhood sessions .Inspired by the Transformer model in machine translation\cite{vaswani2017transformer},we apply a guided-attention mechanism that calculate attention weight for each neighborhood session guided by current session,then the  weighted sum vector  of all neighborhood session representations is used as the complementary collaborative representation.Especially,the attention weights are calculated based on the sequential characteristics and  the temporal recency, since the nearness of sessions in time(recency) has been shown to be useful while determining similar sessions\cite{garg2019sequencetime}.Finally, we design a simple co-attention mechanism to determine the importance of  collaborative information when predicting the next item.Extensive experiments conducted on two real-world  datasets demonstrate the effectiveness of our proposed framework.Our main contributions are listed as follows:

1.we propose a novel NETA model,which combines the ability of  k-nearest-neighbors(KNN) approach to find similar neighborhood sessions effectively and the advantage of neural network to model the sequential behavior within the current session.

2.A novel attention mechasim is proposed,in which the attention weights are calculated by sequential behavior characteristics and  the temporal recency.

3.We conducted extensive experiments on two real-world dataset
 Experimental results show that our proposed model achieves state-of-the-art.

\begin{figure}[ht] 
  \setlength{\abovecaptionskip}{-0.2cm}  
  \setlength{\belowcaptionskip}{-0.7cm} 
    \centering %使图片居中显示
    \includegraphics[width=1\textwidth]{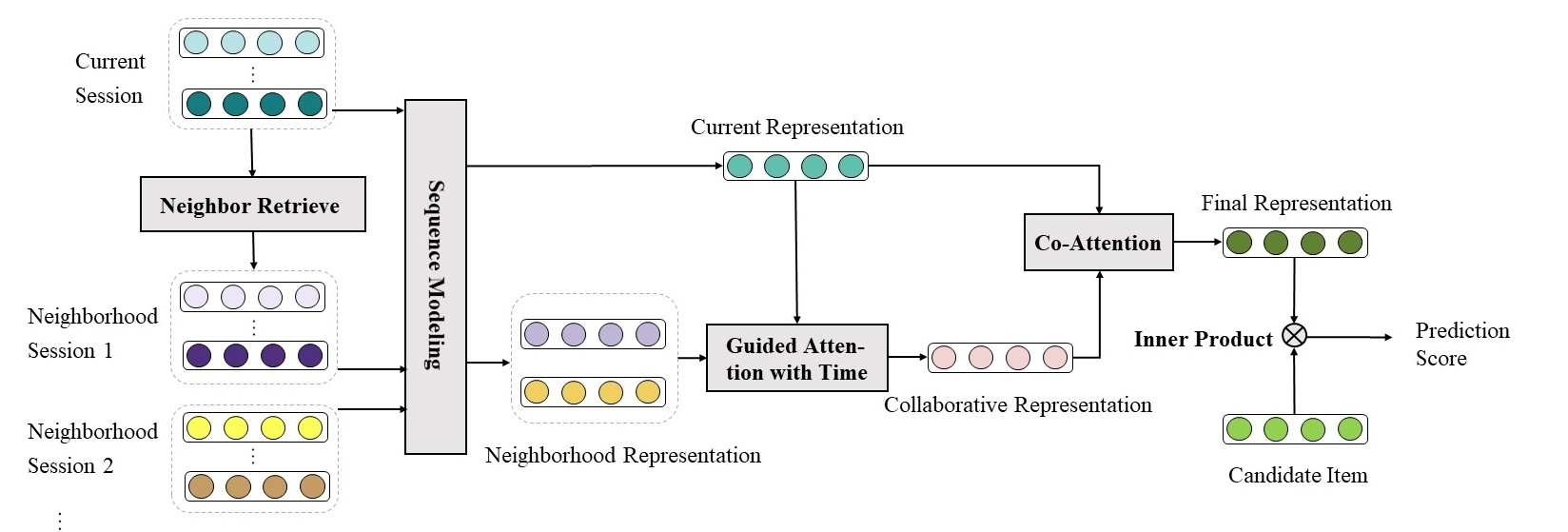} 
    \caption{Pipeline} 
    \label{Pipeline} 
\end{figure}

\section{Methods}
In this section,we first introduce the session based recommendation task.Then we describe the proposed NETA in detail.
\subsection{Session based Recommendation}
Let $V=\left\{v_1,v_2,...,v_m \right\}$ denotes a set of all unique items that appears in all sessions,and call them item dictionary.A session can be regarded as a sequence of click actions created by an anonymous vistor,eg.listening to a song,watching a video.Let $s=\left\{v_{s,1},v_{s,2},...,v_{s,n}\right\}$ denote session $s$,where $v_{s,n}\in V$ denotes a item is clicked at timestamp n in session s.
$s_t=\left\{v_{s,1},v_{s,2},...,v_{s,t} \right\}$ denotes the prefix of session s truncated at t-th timestamp.Our model can generate a ranking list over the item dictionary,and calculate the predicted probability for each item,eg.$\hat
{y}=\left\{\hat{y}_1,\hat{y}_2,...,\hat{y}_m \right\}$,where $\hat{y}_i$ denotes the 
recommendation score for the i-th item in the item dictionary.Finally,the top-$k$ 
items in $\hat{y}$ are recommended.

\subsection{Neighborhood Retrieval}
As shown in Figure 1,the first step is to retrieve neighborhood sessions,which aims to learn dynamic collaborative information for the current session which is consisted of limited actions.Neighborhood sessions are associated with those sessions which also interacted with the same item in current session.The whole process is finished in two steps.

$Step1$:The first step is to search for sessions who interacted with the same item,i.e. like-minded users share similar interests,which are reflected in similar sessions.This step is specially valuable for cold-start scenario.Following SessionKNN\cite{jannach2017knnrnn},we focus on the whole session instead of the last item(eg.ItemKNN) when finding neighborhood sessions.Note that the session's lookup process is implemented by two hash tables that store the map of session-id to item-id and the map of item-id to session-id, which is highly efficient.

$Step2$:The second step is to determine the k most similar neighbors.Technically,given a session $s$,we first determine k most similar past sessions(neighbors) $N_s$ by applying a suitable similarity measure,e.g.,the cosine similarity.To be specific,each session is represented as a binary vector in the m-dimensional space of items(value of 1 for n-th dimension means the n-th item in item dictionary is in this session).The cosine distance between binary vectors reflects quite well the similarity between sessions.Cosine similarity between $s$ and $s_j$ is given as follow:

\begin{center}$sim\left(s,s_j\right)=\frac{\vec s\cdot\vec s_j}{\sqrt{l\left(s \right) \cdot l\left(s_j\right)}}$\end{center}

Now,given the current session $s$,its whole neighbors $N_s$ can be found and we choose the k most similar sessions as neighborhood sessions.

\subsection{Guided attention with time aware}

Since the current session only has limited behaviors and lacks user 's profile or long-term interactions, which is insufficient for a
accurate prediction.In order to improve the precision and diversity of predictions, we make full use of neighborhood sessions by extracting related collaborative information. Thus we propose a novel attention mechanism motivated by Transformer\cite{vaswani2017transformer}.

\begin{figure}[h] 
    \centering %使图片居中显示
    \includegraphics[scale=0.5]{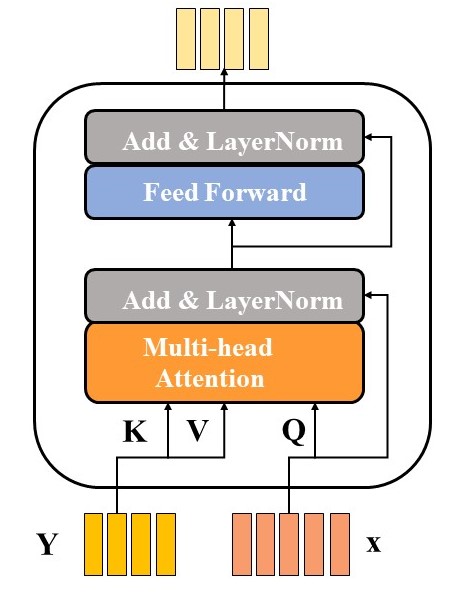}
    \caption{Guided Attention} 
    \label{Guide Attention} 
\end{figure}

Considering giving a query $q \in\mathbb{R}^{1\times d}$,a key matrix  $K \in\mathbb{R}^{n\times d}$ and a value matrix  $V \in\mathbb{R}^{n\times d}$.We calculate each similarity between query and each vector(row) in key,and apply a softmax function to obtain the final attention weights on value.Here the similarity function is "Scaled Dot-Product Attention".The attended feature is the weighted sum over value $V$ based on attention weights.The attended feature $f \in \mathbb{R}^{1\times d}$ is computed as:

\begin{center}$Attention(q,K,V)=softmax(\frac{qK^T}{\sqrt d})V$\end{center}

In order to improve the expressing ability of attention.Mului-head attention is introduced to jointly pay attention to representation from different subspace at different positions,e.g,some heads may focus on long-term interests while others may concentrate on short-term interests.

\begin{center}$MultiHead(q,K,V)=Concat(head_1,...,head_h)W^O$\end{center}

\begin{center}$where\quad head_i=Attention(qW^q_i,KW^K_j,VW^V_j)$\end{center}

Notice that the projection matrices $W^q_i \in\mathbb{R}^{d\times d_h},W^K_i \in\mathbb{R}^{d\times d_h},W^V_i \in\mathbb{R}^{d\times d_h}$ and $W^O \in\mathbb{R}^{h\times d_h\times d}$

Here in our problem,we need to generate collaborative information from neighborhood sessions guided by current session.For example,we can use current session's main purpose as query and regard neighborhood sessions' sequential behaviors as key and value.Thus,the generated collaborative representation for the current session can be understood as reconstructing it by all the neighborhood sessions with respect to their scaled dot-product similarity to current session,i.e generating similar behaviors in neighborhood sessions based on main purpose of current session.

 Recent research work\cite{garg2019sequencetime} shows,nearness of sessions in time(recency) has been shown to be of great use while determining similar neighborhood sessions.Since the occurrences of items in the item dictionary do not obey the assumption of independent and identical distribution(iid assumption),an item will only appear in a session when it is released\cite{tan2016improved}.Besides,some items are timeliness and tend to be clicked 
repeatedly during a certain period, such as seasonal fruits and hot products in e-commerce. Therefore,it is not suitable for current session to consider all neighborhood sessions from different periods as equally significant.

One way to solve this problem is, we choose a head in multi-head attention to focus on the time intervals between neighbor sessions and current session,namely Time Head, which pays more attention to closely sessions.

Since the multi-head attention learns pairwise relationship between current session and each of neighborhood sessions,i.e $<s,N_s\left( i\right)>$,where $N_s\left( i\right)$ denotes the i-th sample of  neighborhood sessions.Time-Head calculates:

\begin{center}$\delta(t)=\phi\left\{
W_tlog\left[t(s)-t(N_s(i))\right]+b\right\}$\end{center}

\begin{center}$TimeHead\left(i,N_s\left( i\right)\right)=NormHead\times  W_{alpha}\delta(t)$\end{center}

where $t\left(s\right)$ is the occurrence time of current session.Time-Head mechanism emphasizes temporal recency by adding the weights of time intervals to multi-head attention.

\subsection{Sequence modeling}
We proceed to present two sequential modeling methods by applying our proposed framework.Two sequential models for SBS are used as examples in this paper,i.e.,NARM\cite{li2017neural} and STAMP\cite{liu2018stamp}.We denote them as NETA-NARM and NETA-STAMP repectively.

NARM consists of two components:a global encoder and a local encoder.The global encoder focuses on modeling user's sequential behavior and the local encoder aims at capture user's main intention in the current session.First the GRU layer converts the input action sequence $v =\left[v_{s,1},v_{s,2},...,v_{s,t} \right]$ into s set to high-dimensional hidden representation $h =\left[h_1,h_2,...,h_t \right]$.Then the final hidden state $h_t$ is used as sequential behavior representation,i.e.$c^{global}_t=h_t$.The weighted sum of all items can reflect user's main purpose,$c^{local}_t=\sum^t_{j=1}\alpha_{tj}h_j$ i.e.which kind of items should pay more attention to.

STAMP model aims to capture user's long-term and short-term interests in order to obtain user's main purpose.Long-term interests is generated by an attention mechanism over items in the current session.Short-term interest is simply defined as the embedding of the last clicked item $v_{s,t}$.Then two fully-connected layers are used for feature abstraction,$c^{long}_t$ and $c^{short}_t$ indicate user's long-term interest and short-term interest finally.

Here NARM and STAMP are used as a sequence feature extractor to transform item embeddings of a session into a session representation.Take NETA-NARM as an example,we apply NARM for both the current session and neighborhood sessions,and extract the collaborative information from sequential behaviors of neighborhood sessions guided by the main purpose of current session.

\subsection{Prediction layer}

Both current session representation($c^{local}_t/c^{global}_t$ or $c^{short}_t/c^{long}_t$) and complementary representation $c^{neighbor}_t$ from neighborhood sessions have strengths and weaknesses.It is essential to accommodate them.Instead of concatenating them easily ,we design an adaptive method for information fusion.We apply a co-attention mechanism to current session representation and complementary representation to determine which part should play a more important role.

The final session representation is computed as :

\begin{center}$c_t = \left[c^{local}_t ,c^{global}_t\right]W_{lg} + \alpha c^{neighbor}_tW_n$\end{center}

where $\alpha = \sigma\left(W_{\alpha l}c^{local}_t+W_{\alpha g}c^{global}_t+W_{\alpha n}c^{neighbor}_t+b \right) $.
$W_{lg} \in\mathbb{R}^{2d\times d}$`, `$W_n,W_{\alpha l},W_{\alpha g},W_{\alpha n} \in\mathbb{R}^{d\times d}$.

Let $V_i$ be the i-th item in item dictionary and here can be regarded as a single candidate item.We generate the final recommendation score by calculating dot product of each candidate item and the final session representation $c_t$:

\begin{center}$\hat z_i = V^T_ic_t$\end{center}

Predicting the next item is essentially a matter of forcing the model to predict the embedding of next item.For example,if you click on three bottles of milk and a dozen eggs,then the final session representation will lie between the milk and the egg,the model will recommender some items which are closely to the embeddings of the milk and the egg in the semantic space.

The objective function is defined as the cross-entropy of the prediction and ground truth.

\begin{center}
    $\mathcal L \left(\hat y \right) = - \sum^m_{i=1}y_ilog\left(\hat y_i \right)+ \left(1-y_i \right)log\left(1-\hat y_i \right)$
\end{center}

where $\hat y=softmax\left(\hat z\right)$ and $y$ denotes the one-hot vector of the ground truth item.

\section{Experimental Setup}
\subsection{Research Questions}
In this section,we first detail our experimental setup.And then we conduct experiments to evaluate the performance of our NETA model to answer the following research questions:

(RQ1)What is the performance of NETA in session-based recommendation tasks?Does it achieve state-of-the-art performance.

(RQ2)How does the collaborative information influence the performance of NETA model?

(RQ3)How does the temporal recency influence the performance of NETA model?

(RQ4)How do the key hyper-parameters influence the performance of NETA model,such as the number of neighbors and the weight of the Time Head?

\subsection{Datasets}
We conduct all the experiments on two real-world datasets:Diginetica and Retailrocket.Diginetica comes from CIKM Cup 2016,among them we only used the released transactional data.Retailrocket comes from an e-commerce personalization company,which contains six months of user browsing activities.Following\cite{ludewig2018evaluation},we manually divide the interaction history into sessions through a 30-minute interval.Clicks within 30 minutes of the same user will be regarded as being from the same session,which is consistent with the server. Following \cite{tan2016improved,li2017neural,liu2018stamp},we filter out sessions of length 1 and items that appear less than 5 times or only appear in test set.Testing set consists of the sessions from subsequent week. After the pre-processing phase,there remains 202633 sessions of 982961 clicks on 43097 items in Diginetica dataset,and 113433 sessions of 413648 clicks on 24095 items in Retailrocket dataset.

Following\cite{tan2016improved,li2017neural,liu2018stamp},for a input session $S=\left\{s_1,s_2,...,s_n\right\}$,we generate the sequences and corresponding labels $\left(\left[s_1\right],s_2 \right),\left(\left[s_1,s_2\right],s_3\right),\left(\left[s_1,s_2,...,s_{n-1}\right],s_n\right)$ for training set and testing set.The statics of the two datasets are shown in Table 1.

\subsection{Experimental Settings}
{\bfseries Baselines.}We use the following baseline methods for comparison,including state-of-the-art and closely related work.

Pop\cite{hidasi2015session}:Predictor always recommends the most popular items in the training set of the platform ,which is a simple but highly competitive method and commonly used in the recommendation system.

Session-Pop\cite{hidasi2015session}:Predicter recommends popular items based on occurrence frequency of the current session and the whole training set.

ItemKNN\cite{davidson2010ItemKNN}:An item-based k-nearest neighbor method,which recommends the most similar k items for the last item in the current session. Similarity between items is consistent with the co-occurences of two items in the same sessions.

SessionKNN\cite{jannach2017knnrnn}:A session-based k-nearest neighbor method.Scores of candidate items are computed by their occurrences in the neighborhood session when predicting the next item for current session.

GRU4Rec\cite{hidasi2015session}:A RNN-based deep learning model for session based recommendation,which uses a session-parallel mini-batch training process and applys a negative sampling technique during the training phase.

NARM\cite{li2017neural}:An improved model based on GRU4Rec,which consists of a global encoder to capture user's main purpose and a local encoder to model the user's sequential behavior.

STAMP\cite{liu2018stamp}:An novel short-term attention priority model,which considers both the short-term and long-term interests of the current session.It abandons RNN structure and is based on attention mechanism,which is computationally efficient.

CSRM\cite{wang2019collaborative}:A state-of-the-art deep learning model for session based recommendation,which consists of an Inner Memory Encoder and an Outer Memory Encoder,where the IME is for current session modeling and OME remembers the most recent sessions to generate collaborative information.

\begin{table}[htbp]
	\centering  
	\caption{Statistics of the experiment datasets}  
	\label{table1} 
	\begin{tabular}{|c|c|c|c|c|c|} 
		\hline  % 表格的横线
		Dataset&train sessions&test sessions&clicks&items&avg.len\\  
		\hline
	    Diginetica&719470&60858&982961&43097&5.12 \\
		\hline
		Retailrocket&264453&35762&  413648&24095&6.68 \\
	    \hline
	\end{tabular}
\end{table}

{\bfseries Evaluation Metric.}
In order to measure the performance of the SRS models, we apply the following evaluation metrics, which are widely used in related work.

Recall@20:Recall@K indicates the proportion of test samples with the correct recommended items in the top-k position of the ranking list,which is defined as:

\begin{center}$Recall@K=\frac{n_{hit}}{N}$\end{center}

MRR@20:We also use Mean Peciprocal Rank,which is the average of reciprocal ranks of the desie item.

\begin{center}$MRR@K=\frac{1}{N}\sum_{t\in G}\frac{1}{Rank\left(t\right)}$\end{center}

It is notable that main difference between mrr and recall is that the order of the recommended items is considered,which is valuable because the order of recommendations indicates its performance.

{\bfseries Parameters.} 
We implement NETA with Tensorflow and conduct experiments on a GeForce GTX TitanXGPU.

Following\cite{li2017neural,wang2019collaborative},to alleviate over-fitting,we use two dropout layers,the first is right after the embedding layer with 25\% dropout,the second is right before the inner product between session representation and embedding of candidate items with 50\% dropout.

We use 10\% of the training set as validation set for adjustment of hyperparameters for all models that contain hyperparameters.We report the best models which are selected by early stopping based on the Recall@20 score on the validation set.Notice that the validation set does not participate in training the neural networks.

According to the validation set,we use the following hyper-parameters:$\left\{d:100,\eta:0.0005,\lambda:0.9\right\}$,where $d$ is embedding dimension,$\eta$ is learning rate,$ \lambda$ is learning rate decay.The mini-batch settings are $\left\{batchsize:512,epoch:30\right\}$.The number of neighborhood sessions of NETA is selected in $\left\{10,20,30,40,50\right\}$,and finally set to 40.

\section{Results and Analysis}

\subsection{RQ1}

\begin{table}[htbp]
	\centering  
	\caption{Performance Comparison(More experiments need to be conducted for hyperparameters adjustment)}  
	\label{table1} 
	\begin{tabular}{|c|c|c|c|c|} 
		\hline  % 表格的横线
        Dataset&\multicolumn{2}{|c|}{Diginetica} & \multicolumn{2}{|c|}{RetailRocket}\\
		\hline
	   Measures&Recall@20&MRR@20 &Recall@20&MRR@20 \\
	   \hline
	  Pop&0.96&0.24&1.24&0.32\\
     \hline
      Session-Pop&21.11&14.60&40.48&32.04\\
      \hline
      IKNN&21.11&14.60&40.48&
      32.04\\
     \hline
     SKNN&49.79&18.59&61.78
      &34.39\\
      \hline
      GRU4Rec&57.95&24.93&-&- \\
      \hline
     STAMP&62.03&27.38&61.08
      &33.10\\
     \hline
     NETA-STAMP&63.59&28.13&63.22
      &33.64\\
     \hline
     NARM&62.58&27.35&61.79
       &34.07\\
      \hline
     CSRM(NARM)&63.07&27.45&63.64 
       &34.76\\
       \hline
     NETA-NARM&63.80&28.08&64.06
       &34.94\\
       \hline
	\end{tabular}
\end{table}

\begin{table}[htbp]
	\centering  
	\caption{Performance Comparison}  
	\label{table1} 
	\begin{tabular}{|c|c|c|c|c|c|c|c|c|} 
		\hline  % 表格的横线
        Dataset&\multicolumn{4}{|c|}{Diginetica} & \multicolumn{4}{|c|}{RetailRocket}\\
		\hline
	   Measures&Recall@10&MRR@10 &Recall@5&MRR@5&Recall@10&MRR@10 &Recall@5&MRR@5 \\
	   \hline
       STAMP&52.07&26.69&41.04&25.21&53.26&32.57&44.82&31.44 \\
  	   \hline NETA-STAMP&53.44&27.42&42.18&25.9&55.31&33.08&46.13&31.85\\
        \hline
	   NARM&51.91&26.53&40.67&25.02&54.28&33.54&46.09&32.44 \\
	   \hline
       CSRM&52.54&26..98&41.27&25.46&55.86&34.21&46.99&33.03 \\
         \hline
       NETA-NARM&53.02&27.33&41.98&25.85&56.16&34.39&47.37&33.20 \\
        \hline
	\end{tabular}
\end{table}

We compare our NETA model to all baselines and find that NETA achieves state-of-the-art performance. The experimental results are shown in Table 2.

We have the following observations:

1)The session-Pop has substantial gain over the Pop while the main difference them is that Session-Pop repeatedly recommends some items of the current session,which indicates us repeat consumption is of great significance.This is mainly because the session is a collection of similar items to a certain extent, and a user's interest within an individual session can be considered to be stable.

2)As for two KNN-based methods(ItemKNN and SessionKNN),we observe that SessionKNN outperforms ItemKNN.The reason is that SessionKNN makes full use of each click item within the session while ItemKNN only pays attention to the last item,which is obviously insufficient.It is notable that although SessionKNN takes advantage of the entire session and considers collaborative filtering,it neglects the sequential order within the session ,which is exactly the problem solved in this paper.

3)NETA outperforms both conventional baselines and state-of-the-art methods.As for NETA and CSRM,the improvement mean that explicitly finding neighborhood sessions instead of remembering recent sessions is useful,because the latter model may contain a lot of noise from irrelevant sessions.

4)In order to verify the performance of NETA in more realistic scenarios,where recommendation system only gives a few items at once since viewers are impatient.We additionally test the performance of recall@10, recall@5, mrr@10 and mrr@5 and the experimental results are summarized in Table 3.It can be observed that NETA still retains certain advantages,which indicates that NETA tends to make more precise recommendations.

5)In the RetailRocket dataset, little performance difference between the NETA and SessionKNN may indicates us that the sequential order within session is not as important as we assumed in some cases,which is also the direction of our future work.We suppose this is due to different characteristics of different datasets.

\subsection{RQ2}
To illustrate the effectiveness of collaborative information in detail,we compare NARM with NETA-NARM, as well as STAMP with NETA-STAMP.The main difference is that NETA considers precise neighborhood sessions as complementary representations in addition to sequence modeling(NARM/STAMP).NETA-NARM obtains obvious improvements over NARM,and NETA-STAMP obtains apparent improvements over STAMP at the same time. The results prove that considering the sequential characteristics only is insufficient to predict the next-item.It is necessary to involve collaborative information since similar sessions tend to click on similar items. In particular, session is a short-term click sequence in which interests tend to be stable.The collaborative information in this case is more useful than it in long-term personalized recommendation systems.

\subsection{RQ3}
We design two models to verify the validity of applying the time head on the basis of guide attention,which emphasizes the temporal information by taking into account the time intervals between sessions.We take Diginetica as an example and the experimental results are summarized in Table4.
1.The NETA model proposed in the paper.
2.NE refers to NETA without time head.
The results prove that applying the time head positively contributes to the accuracy in predicting the next-item.We observe that NETA outperforms NE. The primary reason for it is that seasonal and popular items may be clicked within a short period of time frequently.Our model considers both the sequential characteristics and time intervals when calculating the similarity between the neighborhood session and the current session,which is proved to be efficient and essential.

\begin{table}[htbp]
	\centering  
	\caption{Performance of Time Head}  
	\label{table1} 
	\begin{tabular}{|c|c|c|c|c|c|c|} 
		\hline  % 表格的横线
       Diginetica Dataset&Recall@20&Mrr@20
       &Recall@10&Mrr@10&Recall@5&Mrr@5 \\
       \hline
        NETA-NARM&63.80&28.08&53.02&27.33&41.98&25.85\\
       \hline
        NE-NARM&63.59&27.85&53.18&27.12&41.64&25.57\\
         \hline
        NETA-STAMP&63.59&28.13&53.44&27.42&42.18&25.91\\
         \hline
        NE-STAMP&63.34&27.92&53.11&27.20&41.98&25.72\\
        \hline
	\end{tabular}
\end{table}

\section{Conclusion}
In this paper,a novel framework named Neighborhood-Enhanced and Time-Aware Model(NETA) for session-based recommendation is proposed.NETA combines the ability of  k-nearest-neighbors(KNN) approach to find similar neighborhood sessions effectively and the advantage of neural network to model the sequential behavior.Additionally,NETA considers the temporal recency between sessions,which is proved to be efficient.Extensive experimental results have shown that our model outperformed the state-of-the-art baseline methods.

\bibliographystyle{ieeetr}  
\bibliography{references}
\end{document}